# DeepInf: Social Influence Prediction with Deep Learning

Jiezhong Qiu[†], Jian Tang[♯♭], Hao Ma[‡], Yuxiao Dong[‡], Kuansan Wang[‡], and Jie Tang[†*]

[†]Department of Computer Science and Technology, Tsinghua University
[‡]Microsoft Research, Redmond
[♯]HEC Montreal, Canada
[♭]Montreal Institute for Learning Algorithms, Canada
qiujz16@mails.tsinghua.edu.cn,jian.tang@hec.ca
{haoma,yuxdong,kuansanw}@microsoft.com,jietang@tsinghua.edu.cn

## ABSTRACT

Social and information networking activities such as on Facebook, Twitter, WeChat, and Weibo have become an indispensable part of our everyday life, where we can easily access friends' behaviors and are in turn influenced by them. Consequently, an effective social influence prediction for each user is critical for a variety of applications such as online recommendation and advertising.

Conventional social influence prediction approaches typically design various hand-crafted rules to extract user- and network-specific features. However, their effectiveness heavily relies on the knowledge of domain experts. As a result, it is usually difficult to generalize them into different domains. Inspired by the recent success of deep neural networks in a wide range of computing applications, we design an end-to-end framework, DeepInf[1], to learn users' latent feature representation for predicting social influence. In general, DeepInf takes a user's local network as the input to a graph neural network for learning her latent social representation. We design strategies to incorporate both network structures and user-specific features into convolutional neural and attention networks. Extensive experiments on Open Academic Graph, Twitter, Weibo, and Digg, representing different types of social and information networks, demonstrate that the proposed end-to-end model, DeepInf, significantly outperforms traditional feature engineering-based approaches, suggesting the effectiveness of representation learning for social applications.

[*]Jie Tang is the corresponding author.
[1]Code is available at https://github.com/xptree/DeepInf.

## 1 INTRODUCTION

Social influence is everywhere around us, not only in our daily physical life but also on the virtual Web space. The term social influence typically refers to the phenomenon that a person's emotions, opinions, or behaviors are affected by others. With the global penetration of online and mobile social platforms, people have witnessed the impact of social influence in every field, such as presidential elections [7], advertising [3, 24], and innovation adoption [42]. To date, there is little doubt that social influence has become a prevalent, yet complex force that drives our social decisions, making a clear need for methodologies to characterize, understand, and quantify the underlying mechanisms and dynamics of social influence.

Indeed, extensive work has been done on social influence prediction in the literature [26, 32, 42, 43]. For example, Matsubara et al. [32] studied the dynamics of social influence by carefully designing differential equations extended from the classic 'Susceptible-Infected' (SI) model; Most recently, Li et al. [26] proposed an end-to-end predictor for inferring cascade size by incorporating recurrent neural network (RNN) and representation learning. All these approaches mainly aim to predict the global or aggregated patterns of social influence such as the cascade size within a time-frame. However, in many online applications such as advertising and recommendation, it is critical to effectively predict the social influence for each individual, i.e., user-level social influence prediction.

In this paper, we focus on the prediction of user-level social influence. We aim to predict the action status of a user given the action statuses of her near neighbors and her local structural information. For example, in Figure 1, for the central user $v$, if some of her friends (black circles) bought a product, will she buy the same product in the future? The problem mentioned above is prevalent in real-world applications whereas its complexity and non-linearity have frequently been observed, such as the "S-shaped" curve in [2] and the celebrated "structural diversity" in [46]. The above observations inspire a lot of user-level influence prediction models, most of which [27, 53, 54] consider complicated hand-crafted features, which require extensive knowledge of specific domains and are usually difficult to generalize to different domains.

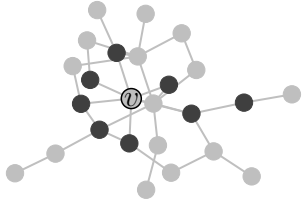

Figure 1: A motivating example of social influence locality prediction. The goal is to predict $v$'s action status, given 1) the observed action statuses (black and gray circles are used to indicate "active" and "inactive", respectively) of her near neighbors and 2) the local network she is embedded in.

Inspired by the recent success of neural networks in representation learning, we design an end-to-end approach to discover hidden and predictive signals in social influence automatically. By architecting network embedding [37], graph convolution [25], and graph attention mechanism [49] into a unified framework, we expect that the end-to-end model can achieve better performance than conventional methods with feature engineering. In specific, we propose a deep learning based framework, DeepInf, to represent both influence dynamics and network structures into a latent space. To predict the action status of a user $v$, we first sample her local neighbors through random walks with restart. After obtaining a local network as shown in Figure 1, we leverage both graph convolution and attention techniques to learn latent predictive signals.

We demonstrate the effectiveness and efficiency of our proposed framework on four social and information networks from different domains—Open Academic Graph (OAG), Digg, Twitter, and Weibo. We compare DeepInf with several conventional methods such as linear models with hand-crafted features [54] as well as the state-of-the-art graph classification model [34]. Experimental results suggest that the DeepInf model can significantly improve the prediction performance, demonstrating the promise of representation learning for social and information network mining tasks.

**Organization** The rest of this paper is organized as follows: Section 2 formulates social influence prediction problem. Section 3 introduces the proposed framework in detail. In Section 4 and 5, we conduct extensive experiments and case studies. Finally, Section 6 summarizes related work and Section 7 concludes this work.

## 2 PROBLEM FORMULATION

In this section, we introduce necessary definitions and then formulate the problem of predicting social influence.

*Definition 2.1.* $r$-**neighbors and** $r$-**ego network** Let $G = (V, E)$ be a static social network, where $V$ denotes the set of users and $E \subseteq V \times V$ denotes the set of relationships[2]. For a user $v$, its $r$-neighbors are defined as $\Gamma_v^r = \{u : d(u, v) \leq r\}$ where $d(u, v)$ is the shortest path distance (in terms of the number of hops) between $u$ and $v$ in the network $G$. The $r$-ego network of user $v$ is the sub-network induced by $\Gamma_v^r$, denoted by $G_v^r$.

*Definition 2.2.* **Social Action** Users in social networks perform social actions, such as retweet. At each timestamp $t$, we observe a binary action status of user $u$, $s_u^t \in \{0, 1\}$, where $s_u^t = 1$ indicates user $u$ has performed this action before or on the timestamp $t$, and $s_u^t = 0$ indicates that the user has not performed this action yet. Such an action log can be available from many social networks, e.g., the "retweet" action in Twitter and the citation action in academic social networks.

Given the above definitions, we introduce social influence locality, which amounts to a kind of closed world assumption: users' social decisions and actions are influenced only by their near neighbors within the network, while external sources are assumed to be not present.

**Problem 1.** *Social Influence Locality[53]* Social influence locality models the probability of $v$'s action status conditioned on her $r$-ego network $G_v^r$ and the action states of her $r$-neighbors. More formally, given $G_v^r$ and $S_v^t = \{s_u^t : u \in \Gamma_v^r \setminus \{v\}\}$, social influence locality aims to quantify the activation probability of $v$ after a given time interval $\Delta t$:

$$P\left(s_v^{t+\Delta t} \middle| G_v^r, S_v^t\right).$$

Practically, suppose we have $N$ instances, each instance is a 3-tuple $(v, a, t)$, where $v$ is a user, $a$ is a social action and $t$ is a timestamp. For such a 3-tuple $(v, a, t)$, we also know $v$'s $r$-ego network—$G_v^r$, the action statuses of $v$'s $r$-neighbors—$S_v^t$, and $v$'s future action status at $t + \Delta t$, i.e., $s_v^{t+\Delta t}$. We then formulate social influence prediction as a binary graph classification problem which can be solved by minimizing the following negative log likelihood objective w.r.t model parameters $\Theta$:

$$\mathcal{L}(\Theta) = -\sum_{i=1}^{N} \log\left(P_\Theta\left(s_{v_i}^{t_i+\Delta t} \middle| G_{v_i}^r, S_{v_i}^{t_i}\right)\right). \quad (1)$$

Especially, in this work, we assume $\Delta t$ is sufficiently large, that is, we want to predict the action status of the ego user $v$ at the end of our observation window.

## 3 MODEL FRAMEWORK

In this section, we formally propose DeepInf, a deep learning based model, to parameterize the probability in Eq. 1 and automatically detect the mechanisms and dynamics of social influence. The framework firstly samples a fixed-size sub-network as the proxy for each $r$-ego network (see Section 3.1). The sampled sub-networks are then fed into a deep neural network with mini-batch learning (see Section 3.2). Finally, the model output is compared with ground truth to minimize the negative log-likelihood loss.

### 3.1 Sampling Near Neighbors

Given a user $v$, a straightforward way to extract her $r$-ego network $G_v^r$ is to perform Breadth-First-Search (BFS) starting from user $v$.

---
[2]In this work, we consider undirected relationships.

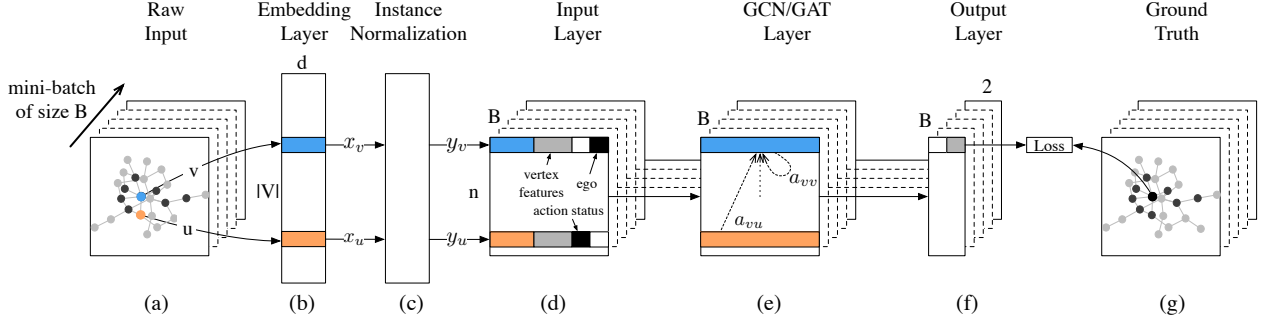

Figure 2: Model Framework of DeepInf. (a) Raw input which consists of a mini-batch of $B$ instances; Each instance is a sub-network comprised of $n$ users who are sampled using random walk with restart as described in Section 3.1. In this example, we keep our eyes on ego user $v$ (marked as blue) and one of her active neighbor $u$ (marked as orange). (b) An embedding layer which maps each user to her $D$-dimensional representation; (c) An Instance Normalization layer [47]. For each instance, this layer normalizes users' embedding $x_u$'s according to Eq. 3. The output embedding $y_u$'s have zero mean and unit variance within each instance. (d) The formal input layer which concatenates together network embedding, two dummy features (one indicates whether the user is active, the other indicates whether the user is the ego), and other customized vertex features (see Table 2 for example). (e) A GCN or GAT layer. $a_{vv}$ and $a_{vu}$ indicate the attention coefficients along self-loop $(v, v)$ and edge $(v, u)$, respectively; The value of these attention coefficients can be chosen between Eq. 5 and Eq. 7 according to the choice between GCN and GAT. (f) and (g) Compare model output and ground truth, we get the negative log likelihood loss. In this example, ego user $v$ was finally activated (marked as black).

However, for different users, $G_v^r$'s may have different sizes. Meanwhile, the size (regarding the number of vertices) of $G_v^r$'s can be very large due to the small-world property in social networks [50]. Such variously sized data is unsuited to most deep learning models. To remedy these issues, we sample a fixed-size sub-network from $v$'s $r$-ego network, instead of directly dealing with the $r$-ego network.

A natural choice of the sampling method is to perform random walk with restart (RWR) [45]. Inspired by [2, 46] which suggest that people are more likely to be influenced by active neighbors than inactive ones, we start random walks from either the ego user $v$ or one of her active neighbors randomly. Next, the random walk iteratively travels to its neighborhood with the probability that is proportional to the weight of each edge. Besides, at each step, the walk is assigned a probability to return to the starting node, that is, either the ego user $v$ or one of $v$'s active neighbors. The RWR runs until it successfully collects a fixed number of vertices, denoted by $\bar{\Gamma}_v^r$ with $|\bar{\Gamma}_v^r| = n$. We then regard the sub-network $\bar{G}_v^r$ induced by $\bar{\Gamma}_v^r$ as a proxy of the $r$-ego network $G_v^r$, and denote $\bar{S}_v^t = \{s_u^t : u \in \bar{\Gamma}_v^r \setminus \{v\}\}$ to be the action statuses of $v$'s sampled neighbors. Therefore, we re-define the optimization objective in Eq. 1 to be:

$$\mathcal{L}(\Theta) = -\sum_{i=1}^{N} \log \left( P_\Theta \left( s_{v_i}^{t_i + \Delta t} \middle| \bar{G}_{v_i}^r, \bar{S}_{v_i}^{t_i} \right) \right). \quad (2)$$

## 3.2 Neural Network Model

With the retrieved $\bar{G}_v^r$ and $\bar{S}_v^t$ for each user, we design an effective neural network model to incorporate both the structural properties in $\bar{G}_v^r$ and action statuses in $\bar{S}_v^t$. The output of the neural network model is a hidden representation for the ego user $v$, which is then used to predict her action status—$s_v^{t+\Delta t}$. As shown in Figure 2, the proposed neural network model consists of a network embedding layer, an instance normalization layer, an input layer, several graph convolutional or graph attention layers, and an output layer. In this section, we introduce these layers one by one and build the model step by step.

**Embedding Layer** With the recent emergence of representation learning [5], the network embedding technique has been extensively studied to discover and encode network structural properties into a low-dimensional latent space. More formally, network embedding learns an embedding matrix $X \in \mathbb{R}^{D \times |V|}$, with each column corresponding to the representation of a vertex (user) in the network $G$. In the proposed model, we use a pre-trained embedding layer which maps a user $u$ to her $D$-dimensional representation $x_u \in \mathbb{R}^D$, as shown in Figure 2(b).

**Instance Normalization [47]** Instance normalization is a recently proposed technique in image style transfer [47]. We adopt this technique in our social influence prediction task. As shown in Figure 2(c), for each user $u \in \bar{\Gamma}_v^r$, after retrieving her representation $x_u$ from the embedding layer, the instance normalization $y_u$ is given by

$$y_{ud} = \frac{x_{ud} - \mu_d}{\sqrt{\sigma_d^2 + \epsilon}} \quad (3)$$

for each embedding dimension $d = 1, \cdots, D$, where

$$\mu_d = \frac{1}{n} \sum_{u \in \bar{\Gamma}_v^r} x_{ud}, \quad \sigma_d^2 = \frac{1}{n} \sum_{u \in \bar{\Gamma}_v^r} (x_{ud} - \mu_d)^2 \quad (4)$$

Here $\mu_d$ and $\sigma_d$ are the mean and variance, and $\epsilon$ is a small number for numerical stability. Intuitively, such normalization can remove instance-specific mean and variance, which encourages the downstream model to focus on users' relative positions in latent embedding space rather than their absolute positions. As we will see later in Section 5, instance normalization can help avoid overfitting during training.

**Input Layer** As illustrated in Figure 2(d), the input layer constructs a feature vector for each user. Besides the normalized low-dimensional embedding comes from up-stream instance normalization layer, it also considers two binary variables. The first variable indicates users' action statuses, and the other indicates whether the user is the ego user. Also, the input layer covers all other customized vertex features such as structural features, content features, and demographic features.

**GCN [25] Based Network Encoding** Graph Convolutional Network (GCN) is a semi-supervised learning algorithm for graph-structured data. The GCN model is built by stacking multiple GCN layers. The input to each GCN layer is a vertex feature matrix, $H \in \mathbb{R}^{n \times F}$, where $n$ is the number of vertices, and $F$ is the number of features. Each row of $H$, denoted by $h_i^\top$, is associated with a vertex. Generally speaking, the essence of the GCN layer is a nonlinear transformation that outputs $H' \in \mathbb{R}^{n \times F'}$ as follows:

$$H' = \text{GCN}(H) = g\left(A(G)HW^\top + b\right), \quad (5)$$

where $W \in \mathbb{R}^{F' \times F}, b \in \mathbb{R}^{F'}$ are model parameters, $g$ is a non-linear activation function, $A(G)$ is a $n \times n$ matrix that captures structural information of graph $G$. GCN instantiates $A(G)$ to be a static matrix closely related to the normalized graph Laplaican [10]:

$$A_{\text{GCN}}(G) = D^{-1/2}AD^{-1/2}, \quad (6)$$

where $A$ is the adjacency matrix[3] of $G$, and $D = \text{diag}(A\mathbf{1})$ is the degree matrix.

**Multi-head Graph Attention [49]** Graph Attention (GAT) is a recent proposed technique that introduces the attention mechanism into GCN. GAT defines matrix $A_{\text{GAT}}(G) = [a_{ij}]_{n \times n}$ through a self-attention mechanism. More formally, an attention coefficient $e_{ij}$ is firstly computed by an attention function $\text{attn}: \mathbb{R}^{F'} \times \mathbb{R}^{F'} \to \mathbb{R}$, which measures the importance of vertex $j$ to vertex $i$:

$$e_{ij} = \text{attn}\left(W h_i, W h_j\right).$$

Different from traditional self-attention mechanisms where the attention coefficients between all pairs of instances will be computed, GAT only evaluates $e_{ij}$ for $(i,j) \in E(\tilde{G}_v^r)$ or $i = j$, i.e., $(i,j)$ is either an edge or a self-loop. In doing so, it is able to better leverage and capture the graph structural information. After that, to make coefficients comparable among vertices, a softmax function is adopted to normalize attention coefficients:

$$a_{ij} = \text{softmax}(e_{ij}) = \frac{\exp(e_{ij})}{\sum_{k \in \tilde{\Gamma}_i^1} \exp(e_{ik})}.$$

Following Velickovic et al. [49], the attention function is instantiated with a dot product and a LeakyReLU [31, 51] nonlinearity. For an edge or a self-loop $(i,j)$, the dot product is performed between parameter $c$ and the concatenation of the feature vectors of the two end points—$Wh_i$ and $Wh_j$, i.e., $e_{ij} = \text{LeakyReLU}\left(c^\top [Wh_i||Wh_j]\right)$, where the LeakyReLU has negative

---
[3] GCN applies self-loop trick on graph $G$ by adding self-loop on each vertex, i.e., $A \leftarrow A + I$

**Table 1: Summary of datasets. $|V|$ and $|E|$ indicates the number of vertices and edges in graph $G = (V,E)$, while $N$ is the number of social influence locality instances (observations) as described in Section 2.**

|  | OAG | Digg | Twitter | Weibo |
|---|---|---|---|---|
| $|V|$ | 953,675 | 279,630 | 456,626 | 1,776,950 |
| $|E|$ | 4,151,463 | 1,548,126 | 12,508,413 | 308,489,739 |
| $N$ | 499,848 | 24,428 | 499,160 | 779,164 |

slop 0.2. To sum up, the normalized attention coefficients can be expressed as:

$$a_{ij} = \frac{\exp\left(\text{LeakyReLU}\left(c^\top [Wh_i||Wh_j]\right)\right)}{\sum_{k \in \tilde{\Gamma}_i^1} \exp\left(\text{LeakyReLU}\left(c^\top [Wh_i||Wh_k]\right)\right)}, \quad (7)$$

where $||$ denotes the vector concatenation operation.

Once obtained the normalized attention coefficients, i.e., $a_{ij}$'s, we can plugin $A_{\text{GAT}}(G) = [a_{ij}]_{n \times n}$ into Eq. 5. This completes the definition of a single-head graph attention. In addition, we apply multi-head graph attention as suggested by Velickovic et al. [49] and Vaswani et al. [48]. The multi-head attention mechanism performs $K$ independent single attention in parallel, i.e., we have $K$ independent parameters $W_1, \cdots, W_K$ and attention matrix $A_1, \cdots, A_K$. Multi-head attention aggregate the output of $K$ single attention together through an aggregation function:

$$H' = g\left(\text{Aggregate}\left(A_1(G)HW_1^\top, \cdots, A_K(G)HW_K^\top\right) + b\right). \quad (8)$$

We concatenate the outputs of each single-head attention to aggregate them except an average operator for the last layer.

**Output Layer and Loss Function** This layer (see Figure 2(f)) outputs a two-dimension representation for each user, we compare the representation of the ego user with ground truth, and then optimize the log-likelihood loss as described in Eq. 2.

**Mini-batch Learning** When sampling from $r$-ego network, we force the sampled sub-networks to have a fixed size $n$. Benefiting from such homogeneity, we can apply mini-batch learning here for efficient training. As shown in Figure 2(a), in each iteration, we first randomly sample $B$ instances to be a mini-batch. Then we optimize our model w.r.t. the sampled mini-batch. Such method runs much faster than full-batch learning and still introduces enough noise during optimization.

## 4 EXPERIMENT SETUP

We set up our experiments with large-scale real-world datasets to quantitatively evaluate the proposed DeepInf framework.

### 4.1 Datasets

Our experiments are conducted on four social networks from different domains —- OAG, Digg, Twitter, and Weibo. Table 1 lists statistics of the four datasets.

**OAG**[4] OAG (Open Academic Graph) dataset is generated by linking two large academic graphs: Microsoft Academic Graph [15] and AMiner [44]. Similar to the treatment in [13], we choose 20 popular conferences from data mining, information retrieval, machine learning, natural language processing, computer vision, and database research communities [5]. The social network is defined to be the co-author network, and the social action is defined to be citation behaviors — a researcher cites a paper from the above conferences. We are interested in how one's citation behaviors are influenced by her collaborators.

---
[4] www.openacademic.ai/oag/

**Digg [23]** Digg is a news aggregator which allows people to vote web content, a.k.a, story, up or down. The dataset contains data about stories promoted to Digg's front page over a period of a month in 2009. For each story, it contains the list of all Digg users who have voted for the story up to the time of data collection and the time stamp of each vote. The voters' friendship links are also retrieved.

**Twitter [12]** The Twitter dataset was built after monitoring the spreading processes on Twitter before, during and after the announcement of the discovery of a new particle with the features of the elusive Higgs boson on Jul. 4th, 2012. The social network is defined to be the Twitter friendship network, and the social action is defined to be whether a user retweets "Higgs" related tweets.

**Weibo [53, 54]** Weibo[6] is the most popular Chinese microblogging service. The dataset is from [53] and can be downloaded here.[7] The complete dataset contains the directed following networks and tweets (posting logs) of 1,776,950 users between Sep. 28th, 2012 and Oct. 29th, 2012. The social action is defined as retweeting behaviors in Weibo — a user forwards (retweets) a post (tweet).

**Data Preparation** We process the above four datasets following the practice in existing work [53, 54]. More concretely, for a user $v$ who was influenced to perform a social action $a$ at some timestamp $t$, we generate a positive instance. Next, for each neighbor of the influenced user $v$, if she was never observed to be active in our observation window, we create a negative instance. Our target is to distinguish positive instances from negative ones. However, the achieved datasets are facing data imbalance problems in two respects. The first comes from the number of active neighbors. As observed by Zhang et al. [54], structural features become significantly correlated with social influence locality when the ego user has a relatively large number of active neighbors. However, the number of active neighbors is imbalanced in most social influence data sets. For example, in Weibo, around 80% instances only have one active neighbor and the instances with the number of active neighbors $\geq 3$ only occupies 8.57%. Therefore, when we train our model on such imbalanced datasets, the model will be dominated by observations with few active neighbors. To deal with the imbalance issue and show the superiority of our model in capturing local structural information, we filter out observations with few active neighbors. Especially, in each data set, we only consider instances where ego users have $\geq 3$ active neighbors. The second problem comes from label imbalance. For example, in the Weibo dataset, the ratio between negative instances and positive instances is about 300:1. To address this issue, we sample a more balanced dataset with the ratio between negative and positive to be 3:1.

---

[5]KDD, WWW, ICDM, SDM, WSDM, CIKM, SIGIR, NIPS, ICML, AAAI, IJCAI, ACL, EMNLP, CVPR, ICCV, ECCV, MM, SIGMOD, ICDE, and VLDB.
[6]www.weibo.com
[7]http://aminer.org/Influencelocality

Table 2: List of features used in this work.

| Name | Description |
|---|---|
| Vertex | Coreness [4]. |
|  | Pagerank [35]. |
|  | Hub score and authority score [9]. |
|  | Eigenvector Centrality [6]. |
|  | Clustering Coefficient [50]. |
|  | Rarity (reciprocal of ego user's degree) [1]. |
| Embedding | Pre-trained network embedding (DeepWalk [36], 64-dim). |
| Ego | The number/ratio of active neighbors [2]. |
|  | Density of subnetwork induced by active neighbors [46]. |
|  | #Connected components formed by active neighbors [46]. |

### 4.2 Evaluation Metrics

To evaluate our framework quantitatively, we use the following performance metrics:

**Prediction Performance** We evaluate the predictive performance of DeepInf in terms of Area Under Curve (AUC) [8], Precision (Prec.), Recall (Rec.), and F1-Measure (F1).

**Parameter Sensitivity** We analyze several hyper-parameters in our model and test how different hyper-parameter choices can influence prediction performance.

**Case Study** We use case studies to further demonstrate and explain the effectiveness of our proposed framework.

### 4.3 Comparison Methods

We compare DeepInf with several baselines.

**Logistic Regression (LR)** We use logistic regression (LR) to train a classification model. The model considers three categories of features: (1) vertex features for the ego-user; (2) pre-trained network embedding (DeepWalk [36]) for ego-user; (3) hand-crafted ego-network features. The features we used are listed in Table 2.

**Support Vector Machine (SVM) [17]** We also use support vector machine (SVM) with linear kernel as the classification model. The model use the same features as logistic regression (LR).

**PSCN [34]** As we model social influence locality prediction as a graph classification problem, we compare our framework with the state-of-the-art graph classification models, PSCN [34]. For each graph, PSCN selects $w$ vertices according to a user-defined ranking function, e.g., degree and betweenness centrality. Then for each selected vertex, it assembles its top $k$ near neighbors according to breadth-first search order. For each graph, The above process constructs a vertex sequence of length $w \times k$ with $F$ channels, where $F$ is the number of features for each vertex. Finally, PSCN applies 1-dimensional convolutional layers on it.

**DeepInf and its Variants** We implement two variants of DeepInf, denoted by DeepInf-GCN and DeepInf-GAT, respectively. DeepInf-GCN uses graph convolutional layer as building blocks of our framework, i.e., setting $A(G) = D^{-1/2}AD^{-1/2}$ in Eq. 5. DeepInf-GAT uses graph attention as shown in Eq. 7. However, both DeepInf and PSCN accept vertex-level features only. Due to this limitation, we do not use the ego-network features in these two models. Instead, we expect that DeepInf can discover the ego-network features and other predictive signals automatically.

**Hyper-parameter Setting & Implementation Details** As for our framework, DeepInf, we first perform random walk with a restart probability 0.8, and the size of sampled sub-network is set to be 50. For the embedding layer, a 64-dimension network embedding is pre-trained using DeepWalk [36]. Then we choose to use a three-layer GCN or GAT structure for DeepInf, both the first and second GCN/GAT layers contain 128 hidden units, while the third layer (output layer) contains 2 hidden units for binary prediction. Especially, for DeepInf with multi-head graph attention, both the first and second layer consists of $K = 8$ attention heads with each computing 16 hidden units (for a total of $8 \times 16 = 128$ hidden units). For detailed model configuration, we adopt exponential linear units (ELU) [11] as nonlinearity (function $g$ in Eq. 5). All the parameters are initialized with Glorot initialization [18] and trained using the Adagrad [16] optimizer with learning rate 0.1 (0.05 for Digg dataset), weight decay $5e^{-4}$ ($1e^{-3}$ for Digg dataset), and dropout rate 0.2. We use 75%, 12.5%, 12.5% instances for training, validation and test, respectively; the mini-batch size is set to be 1024 across all data sets.

As for PSCN, in our experiments, we find that the recommended betweenness centrality ranking function does not work well in predicting social influence. We turn to use breadth-first search order starting from the ego user as the ranking function. When BFS order is not unique, we break ties by ranking active users first. We select $w = 16$ and $k = 5$ by validation and then apply two 1-dimensional convolutional layers. The first conv layer has 16 output channels, a stride of 5, and a kernel size of 5. The second conv layer has 8 output channels, a stride of 1, and a kernel size of 1. The outputs of the second layer are then fed into a fully-connected layer to predict labels.

Finally, we allow PSCN and DeepInf to run at most 500 epochs over the training data, and the best model was selected by early stopping using loss on the validation sets. We release the code for PSCN and DeepInf used in this work at https://github.com/xptree/DeepInf, both implemented with PyTorch.

## 5 EXPERIMENTAL RESULTS

We compare the prediction performance of all methods across the four datasets in Table 3 and list the relative performance gain in Table 4, where the gain is over the closest baseline. In addition, we compare the variants of DeepInf and list the results in Table 5. We have several interesting observations and insights.

(1) As shown in Figure 3, DeepInf-GAT achieves significantly better performance over baselines in terms of both AUC and F1, demonstrating the effectiveness of our proposed framework. In OAG and Digg, DeepInf-GAT discovers the hidden mechanism

**Table 3: Prediction performance of different methods on the four datasets (%).**

| Data | Model | AUC | Prec. | Rec. | F1 |
|---|---|---|---|---|---|
| OAG | LR | 65.55 | 32.26 | **69.97** | 44.16 |
|  | SVM | 65.48 | 32.17 | 69.82 | 44.04 |
|  | PSCN | 69.16 | 36.45 | 64.64 | 46.61 |
|  | DeepInf-GAT | **71.79** | **40.77** | 60.97 | **48.86** |
| Digg | LR | 84.72 | 56.78 | 73.12 | 63.92 |
|  | SVM | 86.01 | 63.42 | 67.34 | 65.32 |
|  | PSCN | 87.37 | 64.75 | 68.15 | 66.40 |
|  | DeepInf-GAT | **90.65** | **66.82** | **78.49** | **72.19** |
| Twitter | LR | 78.07 | 45.86 | **69.81** | 55.36 |
|  | SVM | 79.42 | **49.12** | 67.31 | 56.79 |
|  | PSCN | 78.74 | 47.36 | 67.29 | 55.59 |
|  | DeepInf-GAT | **80.22** | 48.41 | 69.08 | **56.93** |
| Weibo | LR | 77.10 | 42.34 | 72.88 | 53.56 |
|  | SVM | 77.11 | 43.27 | 70.79 | 53.71 |
|  | PSCN | 81.31 | 47.72 | 71.53 | 57.24 |
|  | DeepInf-GAT | **82.72** | **48.53** | **76.09** | **59.27** |

**Table 4: Relative gain of DeepInf-GAT in terms of AUC against the best baseline.**

| Method | OAG | Digg | Twitter | Weibo |
|---|---|---|---|---|
| LR | 65.66 | 84.72 | 78.07 | 77.10 |
| SVM | 65.48 | 86.01 | 79.42 | 77.11 |
| PSCN | 69.16 | 87.37 | 78.74 | 81.31 |
| DeepInf-GAT | 71.79 | 90.65 | 80.22 | 82.72 |
| Relative Gain | 3.8% | 3.8% | 1.0% | 1.7% |

**Table 5: Prediction performance of variants of DeepInf (%).**

| Data | Model | AUC | Prec. | Rec. | F1 |
|---|---|---|---|---|---|
| OAG | DeepInf-GCN | 63.55 | 30.28 | **74.36** | 43.03 |
|  | DeepInf-GAT | **71.79** | **40.77** | 60.97 | **48.86** |
| Digg | DeepInf-GCN | 84.15 | 58.76 | 67.61 | 62.88 |
|  | DeepInf-GAT | **90.65** | **66.82** | **78.49** | **72.19** |
| Twitter | DeepInf-GCN | 76.60 | 44.31 | 66.74 | 53.26 |
|  | DeepInf-GAT | **80.22** | **48.41** | **69.08** | **56.93** |
| Weibo | DeepInf-GCN | 76.85 | 42.44 | 71.30 | 53.21 |
|  | DeepInf-GAT | **82.72** | **48.53** | **76.09** | **59.27** |

and dynamics of social influence locality, giving us 3.8% relative performance gain w.r.t. AUC.

(2) For PSCN, it selects a subset of vertices according to a user-defined ranking function. As mentioned in Section 4, instead of using betweenness centrality, we propose to use BFS order-based ranking function. Such ranking function can be regarded as a pre-defined graph attention mechanism where the ego user pays much more attention to her active neighbors. PSCN outperform linear predictors such as LR and SVM but does not perform as well as DeepInf-GAT.

(3) An interesting observation is the inferiority of DeepInf-GCN, as shown in Table 5. Previously, we have seen the success of GCN in may label classification tasks [25]. However, in this application, DeepInf-GCN achieves the worst performance over all the methods. We attribute its inferiority to the homophily assumption of GCN—similar vertices are more likely to link with each other than dissimilar ones. Under such assumption, for a specific vertex, GCN computes its hidden representation by taking an unweighted average over its neighbors' representations. However, in our application, the homophily assumption may not be true. By averaging over neighbors, GCN may mix predictive signals with noise. On the other hand, as pointed out by [2, 46], active neighbors are more important than inactive neighbors, which also encourages us to use graph attention which treats neighbors differently.

Table 6: Prediction performance of DeepInf-GAT (%) with/without vertex features as introduced in Table 2.

| Data | Features | AUC | Prec. | Rec. | F1 |
|---|---|---|---|---|---|
| OAG | × | 68.07 | 34.77 | **66.87** | 45.78 |
| | √ | **71.79** | **40.77** | 60.97 | **48.86** |
| Digg | × | 89.39 | **68.52** | 72.85 | 70.62 |
| | √ | **90.65** | 66.82 | **78.49** | **72.19** |
| Twitter | × | 78.30 | 47.24 | 65.36 | 54.84 |
| | √ | **80.22** | **48.41** | **69.08** | **56.93** |
| Weibo | × | 81.47 | 46.90 | 75.02 | 57.71 |
| | √ | **82.72** | **48.53** | **76.09** | **59.27** |

(4) In experiments shown in Table 3, 4, and 5, we still rely on several vertex features, such as page rank score and clustering coefficient. However, we want to avoid using any hand-crafted features and make DeepInf a "pure" end-to-end learning framework. Quite surprisingly, we can still achieve comparable performance (as shown in Table 6), even we do not consider any hand-crafted features except the pre-trained network embedding.

## 5.1 Parameter Analysis

In this section, we investigate how the prediction performance varies with the hyper-parameters in sampling near neighbors and the neural network model. We conduct the parameter analyses on the Weibo dataset unless otherwise stated.

**Return Probability of Random Walk with Restart** When sampling near neighbors, the return probability of random walk with restart (RWR) controls the "shape" of the sampled $r$-ego network. Figure 3(a) shows the prediction performance (in terms of AUC and F1) by varying the return probability from 10% to 90%. As the increasing of return probability, the prediction performance also increases slightly, illustrating the locality pattern of social influence.

**Size of Sampled Networks** Another parameter that controls the sampled $r$-ego network is the size of sampled networks. Figure 3(b) shows the prediction performance (in terms of AUC and F1) by varying the size from 10 to 100. We can observe a slow increase of prediction performance when we sample more near neighbors. This is not surprising because we have more information as the size of sampled networks increases.

**Negative Positive Ratio** As we mentioned in Section. 5, the positive and negative observations are imbalanced in our datasets. To investigate how such imbalance influence the prediction performance, we vary the ratio between negative and positive instances from 1 to 10, and show the performance in Figure 3(c). We can observe a decreasing trend w.r.t. the F1 measure, while the AUC score stays stable.

**#Head for Multi-head Attention** Another hyper-parameter we analyze is the number of heads used for multi-head attention. For a fair comparison, we fixed the number of total hidden units to be 128. We vary the number of heads to be 1, 2, 4, 8, 16, 32, 64, 128, i.e., each head has 128, 64, 32, 16, 8, 4, 2, 1 hidden units, respectively. As shown in Figure 3(d), we can see that DeepInf benefits from the multi-head mechanism. However, as the decreasing of the number of hidden units associated with each head, the prediction performance decreases.

**Effect of Instance Normalization** As claimed in Section 3, we use an Instance Normalization (IN) layer to avoid overfitting, especially when training set is small, e.g., Digg. Figure 4(a) and Figure 4(b) illustrate the training loss and test AUC of DeepInf-GAT on the Digg dataset trained with and without IN layer. We can see that IN significantly avoids overfitting and makes the training process more robust.

## 5.2 Discussion on GAT and Case Study

Besides the concatenation-based attention used in GAT (Eq. 7), we also try other popular attention mechanisms, e.g., the dot product attention or the bilinear attention as summarized in [28]. However, those attention mechanisms do not perform as well as the concatenation-based one. In this section, we introduce the order-preserving property of GAT [49]. Based on the property, we attempt to explain the effectiveness of DeepInf-GAT through case studies.

**Observation 1.** *Order-preserving of Graph Attention* Suppose $(i, j)$, $(i, k)$, $(i', j)$ and $(i', k)$ are either edges or self-loops, and $a_{ij}$, $a_{ik}$, $a_{i'j}$, $a_{i'k}$ are the attention coefficients associated with them. If $a_{ij} > a_{ik}$ then $a_{i'j} > a_{i'k}$.

PROOF. As introduced in Eq. 7, the graph attention coefficient for edge (or self-loop) $(i, j)$ is defined as $a_{ij} = \text{softmax}(e_{ij})$, where

$$e_{ij} = \text{LeakyReLU}\left(c^\top \left[W h_i || W h_j\right]\right).$$

If we rewrite $c^\top = \begin{bmatrix} p^\top & q^\top \end{bmatrix}$, we have

$$e_{ij} = \text{LeakyReLU}\left(p^\top W h_i + q^\top W h_j\right).$$

Due to the strict monotonicity of softmax and LeakyReLU, $a_{ij} > a_{ik}$ implies $q^\top W h_j > q^\top W h_k$. Apply the strict monotonicity of LeakyReLU and softmax again, we get $a_{i'j} > a_{i'k}$. □

The above observation shows the following fact—although each vertex only pay attention to its neighbors in GAT (local attention), the attention coefficients have a global ranking, which is determined by $q^\top W h_j$ only. Thus we can define a score function $\text{score}(j) = q^\top W h_j$. Then each vertex pays attention to its neighbors according to this score function—a higher score function value indicates a higher attention coefficient. Thus, plotting the value of the scoring function can illustrate where are the "popular areas" or "important areas" of the network. Furthermore, multi-head attention provides a multi-view mechanism—for $K$ heads, we have $K$ score functions, $\text{score}_k(j) = q_k^\top W_k h_j$, $k = 1, \cdots, K$, highlighting different areas of the network. To better illustrate this mechanism, we perform a few case studies. As shown in Figure 5, we choose four instances

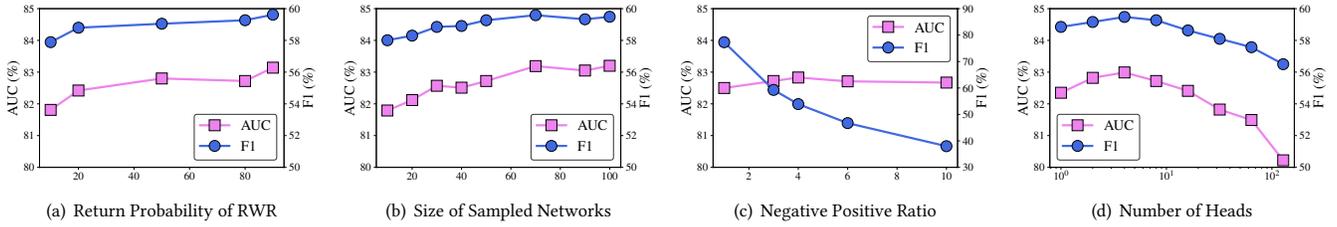

(a) Return Probability of RWR  (b) Size of Sampled Networks  (c) Negative Positive Ratio  (d) Number of Heads

Figure 3: Parameter analysis. (a) Return probability of random walk with restart. (b) Size of sampled networks. (c) Negative positive ratio. (d) The number of heads used for multi-head attention.

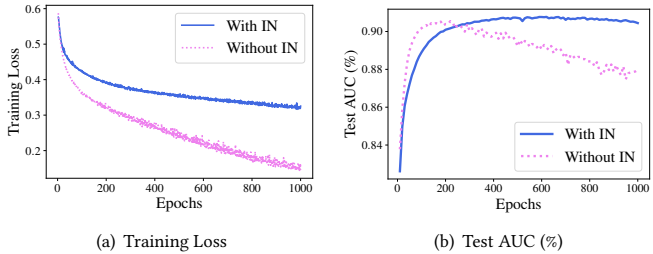

(a) Training Loss  (b) Test AUC (%)

Figure 4: The (a) training loss/(b) test AUC of DeepInf-GAT on Digg data set trained with and without Instance Normalization, vs. the number of epochs. Instance Normalization helps avoid overfitting.

from the Digg dataset (each row corresponding to one instance) and select three representative attention heads from the first GAT layer. Quite interestingly, we can observe explainable and heterogeneous patterns discovered by different attention heads. For example, as shown in Figure 5, the first attention head tend to focus on the ego-user, while the second and the third highlight active users and inactive users, respectively. However, this property does not hold for other attention mechanisms. Due to the page limit, we do not discuss them here.

## 6 RELATED WORK

Our study is closely related to a large body of literature on social influence analysis [42] and graph representation learning [22, 37].

**Social Influence Analysis** Most existing work has focused on social influence modeled as a macro-social process (a.k.a., cascade), with a few that have explored the alternative user-level mechanism that considers the locality of social influence in practice. At the macro level, researchers are interested in global patterns of social influence. Such global patterns includes various respects of a cascade and their correlation with the final cascade size, e.g., the rise-and-fall patterns [32], external influence sources [33], and conformity phenomenon [43]. Recently, there have been efforts to detect those global patterns automatically using deep learning, e.g., the DeepCas model [26] which formulate cascade prediction as a sequence problem and solve it with Recurrent Neural Network.

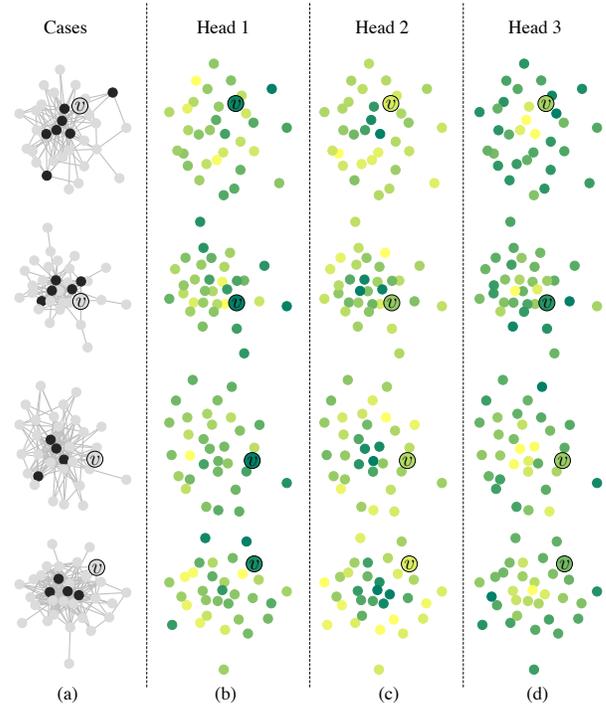

Figure 5: Case study. How different graph attention heads highlight different areas of the network. (a) Four selected cases from the Digg dataset. Active and inactive users are marked as black and gray, respectively. User $v$ is the ego-user that we are interested in. (b)(c)(d) Three representative attention heads.

Another line of studies focuses on the user-level mechanism in social influence where each user is only influenced by her near neighbors. Examples of such work include pairwise influence [19, 39], topic-level influence [42], group formation [2, 38] and structural diversity [14, 29, 46]. Such user-level models act as fundamental building blocks of many real-world problems and applications. For example, in the influence maximization problem [24], both independent cascade and linear threshold models assume a pairwise influence model; In social recommendation [30], a key assumption is social influence—the ratings and reviews of existing users will influence future customers' decisions through social interaction.

Another example is a large-scale field experiment by Facebook Bond et al. [7] during the 2010 US congressional elections, the results showed how online social influence changes offline voting behavior.

**Graph Representation Learning** Representation learning [5] has been a hot topic in research communities. In the context of graph mining, there have been many efforts to graph representation learning. One line of studies focus on vertex (node) embedding, i.e., to learn a low-dimensional latent factors for each vertex. Examples include DeepWalk [36], LINE [41], node2vec [20], metapath2vec [13], NetMF [37], etc. Another line of studies pay attention to representation of graphs, i.e., to learn latent representations of sub-structures for graphs, including, graph kernel [40], deep graph kernel [52], and state-of-the-art method PSCN [34]. Recently, there have been several attempts to incorporate semi-supervised information into graph representation learning. Typical examples include GCN [25], GraphSAGE [21], and the state-of-the-art model GAT [49].

## 7 CONCLUSION

In this work, we study the social influence locality problem. We formulate this problem from a deep learning perspective and propose a graph-based learning framework DeepInf by incorporating the recently developed network embedding, graph convolution, and self-attention techniques. We test the proposed framework on four social and information networks—OAG, Digg, Twitter, and Weibo. Our extensive experimental analysis shows DeepInf significantly outperforms baselines with rich hand-craft features in predicting social influence locality. This work explores the potential of network representation learning in social influence analysis and gives the very first attempt to explain the dynamics of social influence.

The general idea behind the proposed DeepInf can be extended to many network mining tasks. Our DeepInf can effectively and efficiently summarize a local area in a network. Such summarized representations can then be fed into various down-stream applications, such as link prediction, similarity search, network alignment, etc. Therefore, we would like to explore this promising direction for future work. Another exciting direction is the sampling of near neighbors. In this work, we perform random walk with restart without considering any side information. Meanwhile, the sampling procedure is loosely coupled with the neural network model. It is also exciting to combine both sampling and learning together by leveraging reinforcement learning.

**Acknowledgements.** We thank Linjun Zhou, Yutao Zhang, and Jing Zhang for their comments. Jiezhong Qiu and Jie Tang are supported by NSFC 61561130160 and National Basic Research Program of China 2015CB358700.

## REFERENCES
[1] Lada A Adamic and Eytan Adar. 2003. Friends and neighbors on the web. *Social networks* 25, 3 (2003), 211–230.
[2] Lars Backstrom, Dan Huttenlocher, Jon Kleinberg, and Xiangyang Lan. 2006. Group formation in large social networks: membership, growth, and evolution. In *KDD '06*. ACM, 44–54.
[3] Eytan Bakshy, Dean Eckles, Rong Yan, and Itamar Rosenn. 2012. Social influence in social advertising: evidence from field experiments. In *EC '12*. ACM, 146–161.
[4] Vladimir Batagelj and Matjaz Zaversnik. 2003. An O(m) algorithm for cores decomposition of networks. *arXiv preprint cs/0310049* (2003).
[5] Yoshua Bengio, Aaron Courville, and Pascal Vincent. 2013. Representation learning: A review and new perspectives. *IEEE transactions on pattern analysis and machine intelligence* 35, 8 (2013), 1798–1828.
[6] Phillip Bonacich. 1987. Power and centrality: A family of measures. *American journal of sociology* 92, 5 (1987), 1170–1182.
[7] Robert M Bond, Christopher J Fariss, Jason J Jones, Adam DI Kramer, Cameron Marlow, Jaime E Settle, and James H Fowler. 2012. A 61-million-person experiment in social influence and political mobilization. *Nature* 489, 7415 (2012), 295.
[8] Chris Buckley and Ellen M Voorhees. 2004. Retrieval evaluation with incomplete information. In *SIGIR '04*. ACM, 25–32.
[9] Soumen Chakrabati, B Dom, D Gibson, J Kleinberg, S Kumar, P Raghavan, S Rajagopalan, and A Tomkins. 1999. Mining the link structure of the World Wide Web. *IEEE Computer* 32, 8 (1999), 60–67.
[10] Fan RK Chung. 1997. *Spectral graph theory*. Number 92. American Mathematical Soc.
[11] Djork-Arné Clevert, Thomas Unterthiner, and Sepp Hochreiter. 2015. Fast and accurate deep network learning by exponential linear units (elus). *arXiv preprint arXiv:1511.07289* (2015).
[12] Manlio De Domenico, Antonio Lima, Paul Mougel, and Mirco Musolesi. 2013. The anatomy of a scientific rumor. *Scientific reports* 3 (2013), 2980.
[13] Yuxiao Dong, Nitesh V Chawla, and Ananthram Swami. 2017. metapath2vec: Scalable Representation Learning for Heterogeneous Networks. In *KDD '17*. ACM, 135–144.
[14] Yuxiao Dong, Reid A Johnson, Jian Xu, and Nitesh V Chawla. 2017. Structural Diversity and Homophily: A Study Across More Than One Hundred Big Networks. In *KDD '17*. ACM, 807–816.
[15] Yuxiao Dong, Hao Ma, Zhihong Shen, and Kuansan Wang. 2017. A Century of Science: Globalization of Scientific Collaborations, Citations, and Innovations. In *KDD '17*. ACM, 1437–1446.
[16] John Duchi, Elad Hazan, and Yoram Singer. 2011. Adaptive subgradient methods for online learning and stochastic optimization. *JMLR* 12, Jul (2011), 2121–2159.
[17] Rong-En Fan, Kai-Wei Chang, Cho-Jui Hsieh, Xiang-Rui Wang, and Chih-Jen Lin. 2008. LIBLINEAR: A library for large linear classification. *JMLR* 9, Aug (2008), 1871–1874.
[18] Xavier Glorot and Yoshua Bengio. 2010. Understanding the difficulty of training deep feedforward neural networks. In *AISTATS '10*. 249–256.
[19] Amit Goyal, Francesco Bonchi, and Laks VS Lakshmanan. 2010. Learning influence probabilities in social networks. In *WSDM '10*. ACM, 241–250.
[20] Aditya Grover and Jure Leskovec. 2016. node2vec: Scalable feature learning for networks. In *KDD '16*. ACM, 855–864.
[21] Will Hamilton, Zhitao Ying, and Jure Leskovec. 2017. Inductive representation learning on large graphs. In *NIPS '17*. 1025–1035.
[22] William L Hamilton, Rex Ying, and Jure Leskovec. 2017. Representation Learning on Graphs: Methods and Applications. *arXiv preprint arXiv:1709.05584* (2017).
[23] Tad Hogg and Kristina Lerman. 2012. Social dynamics of digg. *EPJ Data Science* 1, 1 (2012), 5.
[24] David Kempe, Jon Kleinberg, and Éva Tardos. 2003. Maximizing the spread of influence through a social network. In *KDD '03*. 137–146.
[25] Thomas N Kipf and Max Welling. 2017. Semi-supervised classification with graph convolutional networks. *ICLR '17* (2017).
[26] Cheng Li, Jiaqi Ma, Xiaoxiao Guo, and Qiaozhu Mei. 2017. DeepCas: An end-to-end predictor of information cascades. In *WWW '17*. 577–586.
[27] Huijie Lin, Jia Jia, Jiezhong Qiu, Yongfeng Zhang, Guangyao Shen, Lexing Xie, Jie Tang, Ling Feng, and Tat-Seng Chua. 2017. Detecting stress based on social interactions in social networks. *TKDE* 29, 9 (2017), 1820–1833.
[28] Minh-Thang Luong, Hieu Pham, and Christopher D Manning. 2015. Effective approaches to attention-based neural machine translation. *EMNLP '15* (2015).
[29] Hao Ma. 2013. An experimental study on implicit social recommendation. In *SIGIR '13*. ACM, 73–82.
[30] Hao Ma, Irwin King, and Michael R Lyu. 2009. Learning to recommend with social trust ensemble. In *SIGIR '09*. ACM, 203–210.
[31] Andrew L Maas, Awni Y Hannun, and Andrew Y Ng. 2013. Rectifier nonlinearities improve neural network acoustic models. In *ICML '13*. 3.
[32] Yasuko Matsubara, Yasushi Sakurai, B Aditya Prakash, Lei Li, and Christos Faloutsos. 2012. Rise and fall patterns of information diffusion: model and implications. In *KDD '12*. 6–14.

[33] Seth A Myers, Chenguang Zhu, and Jure Leskovec. 2012. Information diffusion and external influence in networks. In *KDD '12*. ACM, 33–41.
[34] Mathias Niepert, Mohamed Ahmed, and Konstantin Kutzkov. 2016. Learning convolutional neural networks for graphs. In *ICML '16*. 2014–2023.
[35] Lawrence Page, Sergey Brin, Rajeev Motwani, and Terry Winograd. 1999. *The PageRank citation ranking: Bringing order to the web.* Technical Report. Stanford InfoLab.
[36] Bryan Perozzi, Rami Al-Rfou, and Steven Skiena. 2014. Deepwalk: Online learning of social representations. In *KDD '14*. ACM, 701–710.
[37] Jiezhong Qiu, Yuxiao Dong, Hao Ma, Jian Li, Kuansan Wang, and Jie Tang. 2018. Network Embedding as Matrix Factorization: Unifying DeepWalk, LINE, PTE, and node2vec. In *WSDM '18*. ACM, 459–467.
[38] Jiezhong Qiu, Yixuan Li, Jie Tang, Zheng Lu, Hao Ye, Bo Chen, Qiang Yang, and John E Hopcroft. 2016. The lifecycle and cascade of wechat social messaging groups. In *WWW '16*. 311–320.
[39] Kazumi Saito, Ryohei Nakano, and Masahiro Kimura. 2008. Prediction of information diffusion probabilities for independent cascade model. In *KES '08*. Springer, 67–75.
[40] Nino Shervashidze, SVN Vishwanathan, Tobias Petri, Kurt Mehlhorn, and Karsten Borgwardt. 2009. Efficient graphlet kernels for large graph comparison. In *AISTATS' 09*. 488–495.
[41] Jian Tang, Meng Qu, Mingzhe Wang, Ming Zhang, Jun Yan, and Qiaozhu Mei. 2015. LINE: Large-scale information network embedding. In *WWW '15*. 1067–1077.
[42] Jie Tang, Jimeng Sun, Chi Wang, and Zi Yang. 2009. Social influence analysis in large-scale networks. In *KDD '09*. ACM, 807–816.
[43] Jie Tang, Sen Wu, and Jimeng Sun. 2013. Confluence: Conformity influence in large social networks. In *KDD '13*. ACM, 347–355.
[44] Jie Tang, Jing Zhang, Limin Yao, Juanzi Li, Li Zhang, and Zhong Su. 2008. Arnetminer: extraction and mining of academic social networks. In *KDD '08*. 990–998.
[45] Hanghang Tong, Christos Faloutsos, and Jia-Yu Pan. 2006. Fast Random Walk with Restart and Its Applications. In *ICDM '06*. 613–622.
[46] Johan Ugander, Lars Backstrom, Cameron Marlow, and Jon Kleinberg. 2012. Structural diversity in social contagion. *PNAS* 109, 16 (2012), 5962–5966.
[47] Dmitry Ulyanov, Vedaldi Andrea, and Victor Lempitsky. 2016. Instance Normalization: The Missing Ingredient for Fast Stylization. *arXiv preprint arXiv:1607.08022* (2016).
[48] Ashish Vaswani, Noam Shazeer, Niki Parmar, Jakob Uszkoreit, Llion Jones, Aidan N Gomez, Łukasz Kaiser, and Illia Polosukhin. 2017. Attention is all you need. In *NIPS '17*. 6000–6010.
[49] Petar Velickovic, Guillem Cucurull, Arantxa Casanova, Adriana Romero, Pietro Lio, and Y Bengio. 2018. Graph Attention Networks. *ICLR '18* (2018).
[50] Duncan J Watts and Steven H Strogatz. 1998. Collective dynamics of 'small-world' networks. *nature* 393, 6684 (1998), 440–442.
[51] Bing Xu, Naiyan Wang, Tianqi Chen, and Mu Li. 2015. Empirical evaluation of rectified activations in convolutional network. *arXiv preprint arXiv:1505.00853* (2015).
[52] Pinar Yanardag and SVN Vishwanathan. 2015. Deep graph kernels. In *KDD '15*. 1365–1374.
[53] Jing Zhang, Biao Liu, Jie Tang, Ting Chen, and Juanzi Li. 2013. Social Influence Locality for Modeling Retweeting Behaviors.. In *IJCAI' 13*.
[54] Jing Zhang, Jie Tang, Juanzi Li, Yang Liu, and Chunxiao Xing. 2015. Who influenced you? predicting retweet via social influence locality. *TKDD* 9, 3 (2015), 25.